\documentclass[conference]{IEEEtran}
\IEEEoverridecommandlockouts
\usepackage{cite}
\usepackage{amsmath,amssymb,amsfonts}
\usepackage{algorithmic}
\usepackage{graphicx}
\usepackage{textcomp}
\usepackage{xcolor}
\usepackage{threeparttable}
\usepackage{enumitem}
\def\BibTeX{{\rm B\kern-.05em{\sc i\kern-.025em b}\kern-.08em
    T\kern-.1667em\lower.7ex\hbox{E}\kern-.125emX}}

\setlist[itemize]{leftmargin=*}
\begin{document}

\title{Machine Learning with Real-time and Small Footprint Anomaly Detection System for In-Vehicle Gateway\\
}

\author{
\IEEEauthorblockN{Yi Wang}
\IEEEauthorblockA{\textit{Product Cybersecurity \& Privacy Office}\\
\textit{Continental Automotive Singapore}\\
Singapore\\
estelle.wang@continental.com}
\and
\IEEEauthorblockN{Yuanjin Zheng}
\IEEEauthorblockA{\textit{School of EEE}\\
\textit{Nanyang Technological University}\\
Singapore\\
YJZHENG@ntu.edu.sg}
\and
\IEEEauthorblockN{Yajun Ha}
\IEEEauthorblockA{\textit{School of Information Science and Technology}\\
\textit{ShanghaiTech University}\\
Shanghai, China\\
hayj@shanghaitech.edu.cn}
}

\maketitle

\begin{abstract}
Anomaly Detection System (ADS) is an essential part of a modern gateway Electronic Control Unit (ECU) to detect abnormal behaviors and attacks in vehicles. Among the existing attacks, ``one-time" attack is the most challenging to be detected, together with the strict gateway ECU constraints of both microsecond or even nanosecond level real-time budget and limited footprint of code. To address the challenges, we propose to use the self-information theory to generate values for training and testing models, aiming to achieve real-time detection performance for the ``one-time” attack that has not been well studied in the past. Second, the generation of self-information is based on logarithm calculation, which leads to the smallest footprint to reduce the cost in Gateway. Finally, our proposed method uses an unsupervised model without the need of training data for anomalies or attacks. We have compared different machine learning methods ranging from typical machine learning models to deep learning models, e.g., Hidden Markov Model (HMM), Support Vector Data Description (SVDD), and Long Short Term Memory (LSTM). Experimental results show that our proposed method achieves 8.7 times lower False Positive Rate (FPR), 1.77 times faster testing time, and 4.88 times smaller footprint. 
\end{abstract}


\section{Introduction} \label{section:intro}

With the emerging requirement of the Artificial Intelligence of Things (AIoT) in vehicular applications, consumers pay more and more attention to personal privacy and confidentiality, apart from the safety, compatibility, and performance of a modern vehicle~\cite{Meneghello19}. The in-vehicle network consists of Electronic Control Units (ECUs) to construct various subsystems. In contrast with the point to point connections, several peripherals have been connected using the same set of wires, enabling different controllers to share the same signals of a single sensor. However, the basic protocol of the in-vehicle network, Controller Area Network (CAN) ~\cite{Bosch91}, only provides the address dependability and fault detection.

There are existing security threads and holes to be used by the attackers to retrieve sensitive information of the critical components, such as the engine ECU, the brake ECU, etc.~\cite{Hoppe08} ~\cite{Wolf04}. These lead to possible security vulnerabilities, e.g., spoofing, manipulation, man-in-the-middle attacks. These attacks can be successful when the adversaries compromise the primary interfaces and the gateway system.  For instance, Miller and Valasek~\cite{Miller15} have successfully penetrated the wireless interfaces of the entertainment system in the Jeep Cherokee and took over the central controller of the car. This attack can cause a severe safety concern with a million of losses, which affects all the stakeholders, such as OEMs and automotive tier-one suppliers. There are also existing typical attacks of CAN bus in-vehicle networks, which are message flooding, cyclic message, replay, and ``one-time" attacks. Among them, ``one-time" attack is the most challenging and difficult to be detected by Anomaly Detection System (ADS). This type of attack only manipulates the data payload or the content once using one malicious CAN message (the data payload or content is still within the valid data range after hacking). This kind of attack could be severe for the CAN bus message with sensitive information to critical ECUs, e.g., braking control, air-bag control, and engine control.

To cope with ``one-time" attacks, machine learning based anomaly detection systems have been proposed~\cite{Narayanan15}~\cite{Theissler14}~\cite{Avatefipour19}~\cite{Chockalingam17}~\cite{Wang18}. However, they are facing several challenges. First, the performance of prevention from anomalies highly depends on the response of the Anomaly Detection System (ADS), which requires the microsecond detection response or even the nanosecond detection response in critical automotive applications. Second, the footprint of ADS is required as small as possible to leave more space for other applications on gateway ECU. Third, extra effort is required for labeling dataset in most of the existing machine learning models. In summary, most existing machine learning based ADS are supervised models, which cannot support real-time detection for multiple driving behavior with a small footprint. Till now, there has been only one existing work can efficiently detect ``one-time" attack proposed by M"uter and Asaj~\cite{Muter11}. This work is from Daimler AG (German Car Manufacturer) based on a real-time practical dataset collected from in-vehicle. 

On the other hand, there have existing non-machine learning works about how to detect anomalies without focusing on ``one-time" attacks. Marchetti et al. propose information-theoretic anomaly detection algorithms focused on the detection of replay attacks and random fuzzing attacks~\cite{Marchetti16}. Harma and M{\" {o}}ller prove that an an ADS is required to be applied in the gateway ECU of the vehicle, which is the critical ECU to exchange data packages among other ECUs~\cite{Sharma18}. Katragadda et al. propose a sequence mining approach to detect low-rate injection attacks in CAN, which achieves over 99\% f-score, and outperforms existing dictionary-based and multivariate Markov chain-based approach~\cite{Katragadda20}.


To address these challenges, we propose a novel real-time ADS based on the unsupervised machine learning to successfully prevent the most challenging ``one-time" attack, which also achieves malicious detection response within a microsecond and smaller footprint. Our contributions in the paper are as follows.
\begin{itemize} 
    \item{First, we use self-information from information theory to generate the values to construct the training and testing matrices of the proposed machine learning model. It reaches up to millisecond quantum to detect the ``one-time" attacks in-vehicle network, which has not been well studied in the existing ADS,} 
     \item{Second, the computation of the proposed model is based on the self-information generation (e.g., logarithm), which leads to the smallest footprint.}
    \item {Third, we reduce the extra effort to process the dataset with the proposed unsupervised model compared to the most existing supervised models. Our method can also be trained with a normal training dataset, which does not require the training data with anomalies or attacks.}
\end{itemize}  




Compared to the best among the state-of-the-art models, e.g., Hidden Markov Model (HMM), Support Vector Data Description (SVDD), and Long Short Term Memory (LSTM) based detection, experimental results show that our proposed method achieves 8.7 times lower False Positive Rate (FPR), 1.77 times faster testing time, and 4.88 times smaller footprint.

The reminder of this paper is organized as follows.  Section~\ref{section:ml} proposes real-time ADS based on the lightweight machine learning model with training and testing. Section~\ref{section:result} shows the experimental results of our proposed real-time ADS in-vehicle platform and comparison with the previous works. Finally, Section~\ref{section:conclusion} concludes the paper.

\section{Proposed Real-time ADS based on Machine Learning}
\label{section:ml}
The challenges to providing efficient ADS on Gateway ECU are preventing the challenging attacks (``one-time" and replay attacks) with real-time detection in microseconds or even nanoseconds, and a smaller footprint to lower the cost for ECU. In this section, we first introduce the background of the existing ADS. Then, we propose a lightweight machine learning model with the adaptation of the information theory to have the ability to detect ``one-time" and replay attacks in-vehicle network, which can be easily reused for detect message flooding, and cyclic message attacks.
\subsection{The Workflow Of Existing Anomaly Detection and Prevention Systems}
\begin{figure}
\centering
\includegraphics[width=3.1in]{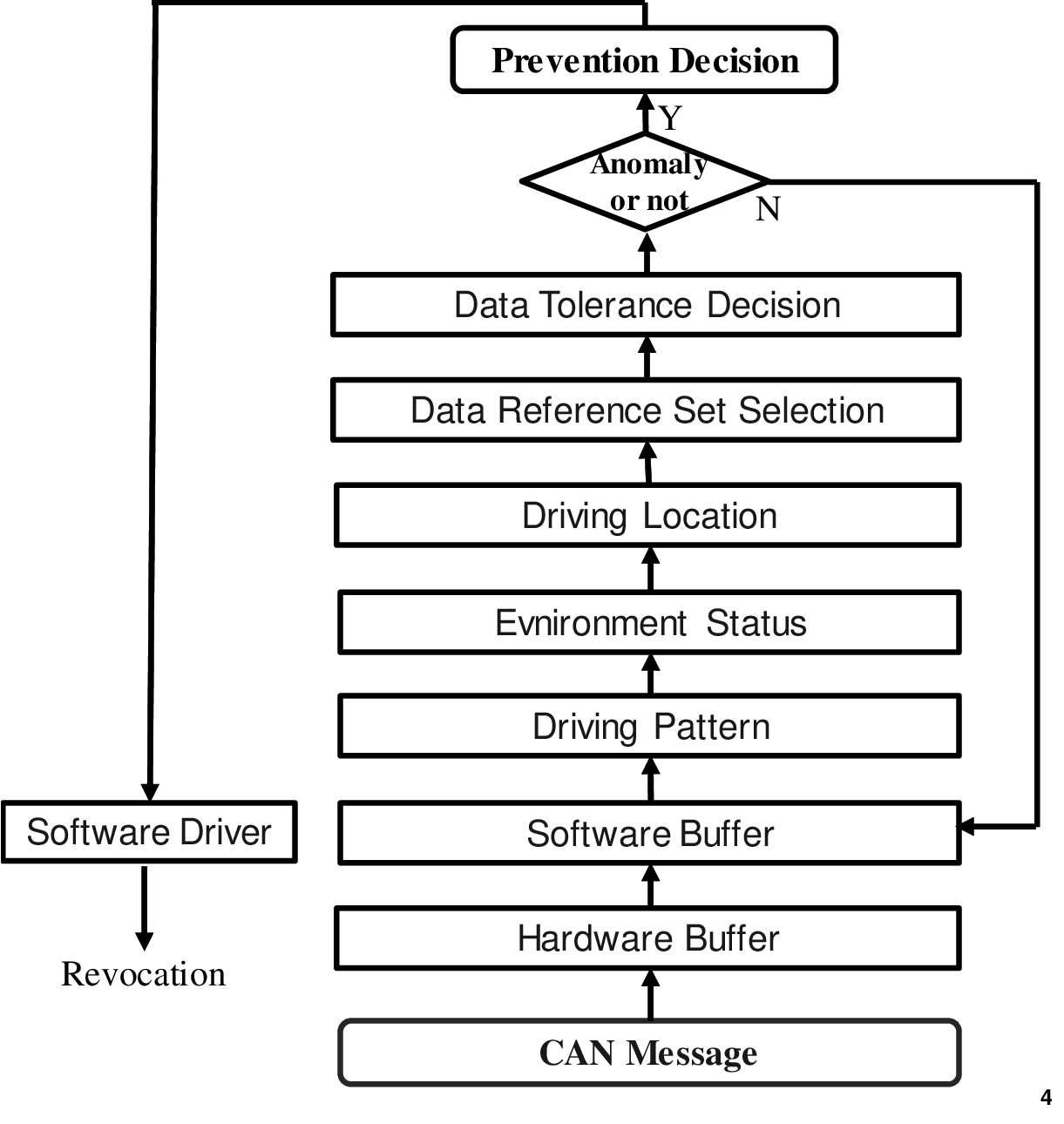}
\caption{The work flow of an existing anomaly detection and prevention system}
\label{fig:adsflow}
\end{figure}

Fig.~\ref{fig:adsflow} shows the workflow of an existing anomaly detection and prevention system. The existing solutions to detect the anomaly in the payload (data) content of the message rely on the value of the data. The existing solutions compare the incoming data with its references, such as driving pattern, environment status, driving location, data reference set selection, and data tolerance decision. It causes a long decision (huge time delay) of the anomaly, which leads to non-real-time prevention. The CAN messages are observed through the buffers (i.e., hardware and software), driving patterns, and scanned periodically to detect anomalies. Consequently, the delay incurred in the workflow will not facilitate real-time anomaly detection.

\subsection{The Proposed Method}

 Our proposed real-time ADS employs a matrix entropy machine learning independent of the data in the message. During the detection, the reference matrix automatically updates the reference value using the machine learning features. Thus, it eliminates the need for having the diﬀerent sets of reference data for each different condition. Updating the reference can be generalized to the dataset's behavior to reduce the ratio of false detection. In this context, the condition can be associated with different drivers or different road conditions. For instance, the velocity dataset at the highway and the urban roads could have different characteristics, which lead to false detection if the references are not updated. Our proposed lightweight machine learning using conditional self-information matrices comprises two phases, the training, and the testing phases, which will be elaborated as follows.

\textbf{Training Phase}

We characterize the data in the conditional self-information matrix, which models the likelihood of each current dataset in accordance with the previous dataset. This training calculates the conditional self-information for every sequence in the data and uses it to detect anomalies. Here, we use the existing velocity dataset as an example to illustrate our method, which can be modiﬁed to any dataset. Velocity dataset is used to train a matrix with consecutive two values at time t and $t-1$, in which the contents are the conditional self-information of the velocity value at the time of $t$ to the velocity value at the time of $t-1$.

\begin{figure}
\centering
\includegraphics[width=3.5in]{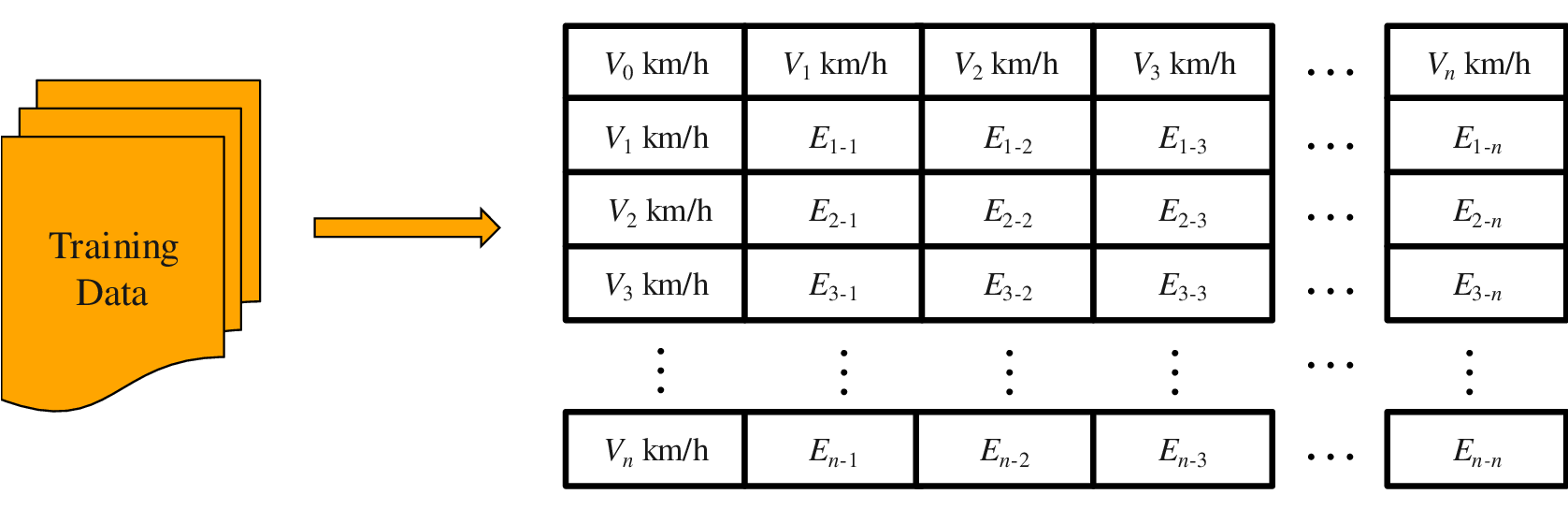}
\caption{Training phase: matrix with conditional self-information}
\label{fig:train}
\end{figure}

Fig.~\ref{fig:train}  shows the structure of the trained matrix where $V_0, V_1, V_2, V_n$ represent the valid values of the existing dataset of velocity, in which $V_0$ is the minimum velocity value in the dataset and $V_n$ is the maximum velocity value in the same dataset. The selection between $V_0$ and $V_n$ must consider the characteristic of the existing dataset. $E_{i-j} (i \in  0, 1, 2, 3, ... , n)$, represents the conditional self-information of a velocity $V_i$ with probability $P(V_i|V_j)$ under the previous occurrence of $V_j$. It can be computed as follows:
\begin{equation}
E_{i-j} =\log_2 \frac{1}{P(V_i|V_j)}
\label{equation:new}
\end{equation}
In the training phase, the sequences of the vehicular velocity gathered from the test vehicles are analyzed in order to model the normal velocity of the ﬂuctuation behaviors of a vehicle. The conditional self-information of each velocity, in accordance to the previous velocity, is calculated and inserted into a two-dimensional matrix of order $n$, where $n$ is the highest possible velocity value reached by the vehicle (i.e., 250 km/h). Upon the training, the trained matrix will be stored in the gateway ECU as a reference look-up-table (LUT) for the future detection.

\textbf{Testing Phase}

\begin{figure}
\centering
\includegraphics[width=3.5in]{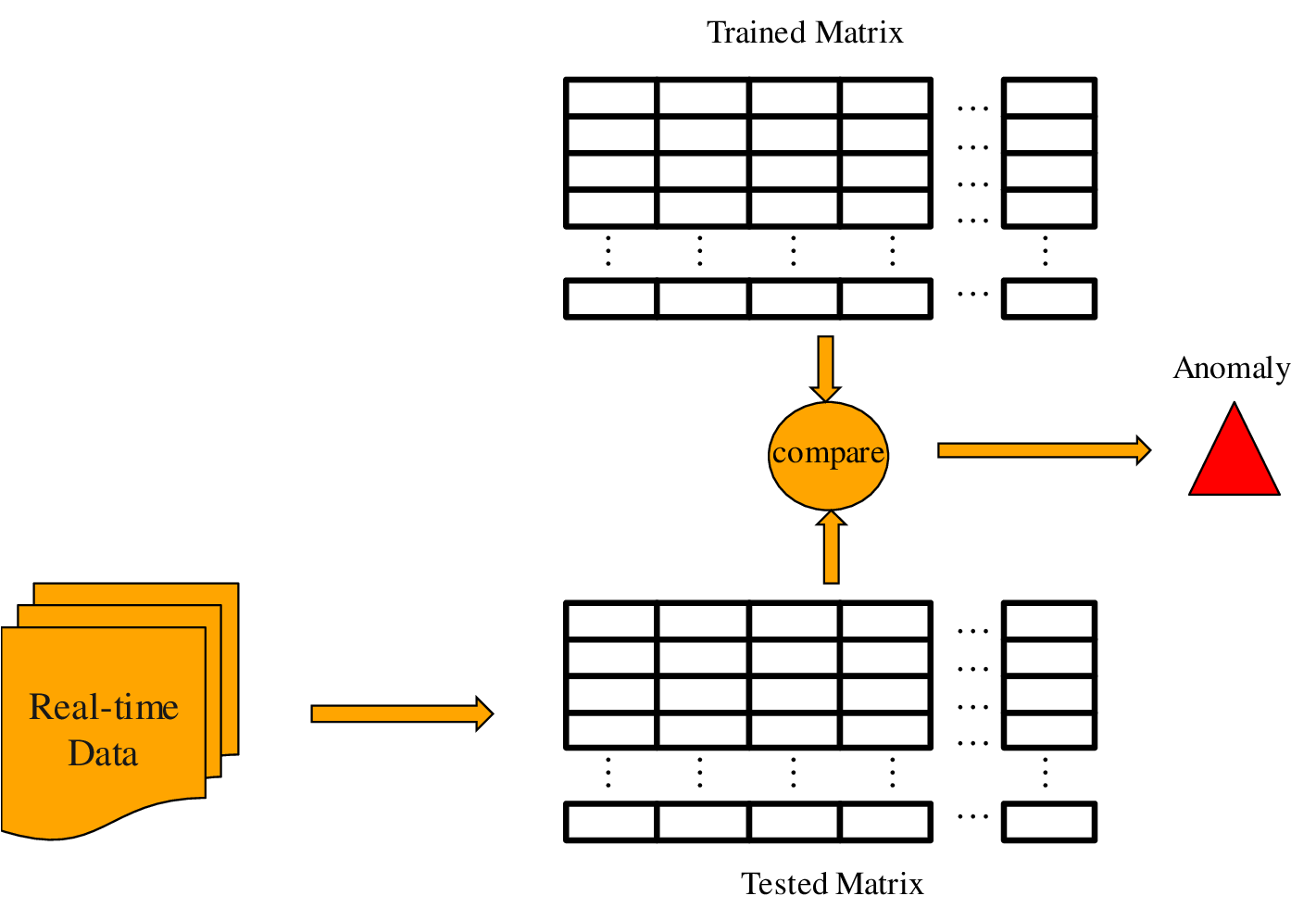}
\caption{Testing phase: anomaly is detected by comparing the trained matrix with the tested matrix}
\label{fig:test}
\end{figure}

The testing phase initiates similarly to the training phase. A conditional self-information matrix is created using the testing dataset. By the use of both the training matrix and the testing matrix, final anomaly detection is achieved by comparing the difference of the conditional self-information between these two matrices. Fig.~\ref{fig:test} depicts the real-time detection procedure and how to determine the anomaly. 

The tested matrix can be scaled to ﬁt the critical time requirements of the gateway ECU. Since both the trained matrix and the tested matrix have the same ﬁxed sizes, it will not increase the occupied size with the future upgrading requirements. This method would be able to detect the most common anomalies, e.g., spooﬁng, the bad injection attacks, and the replay attacks.

\section{Experimental Results}
\label{section:result}





\begin{figure}[t]
\centering
\includegraphics[width=3.5in]{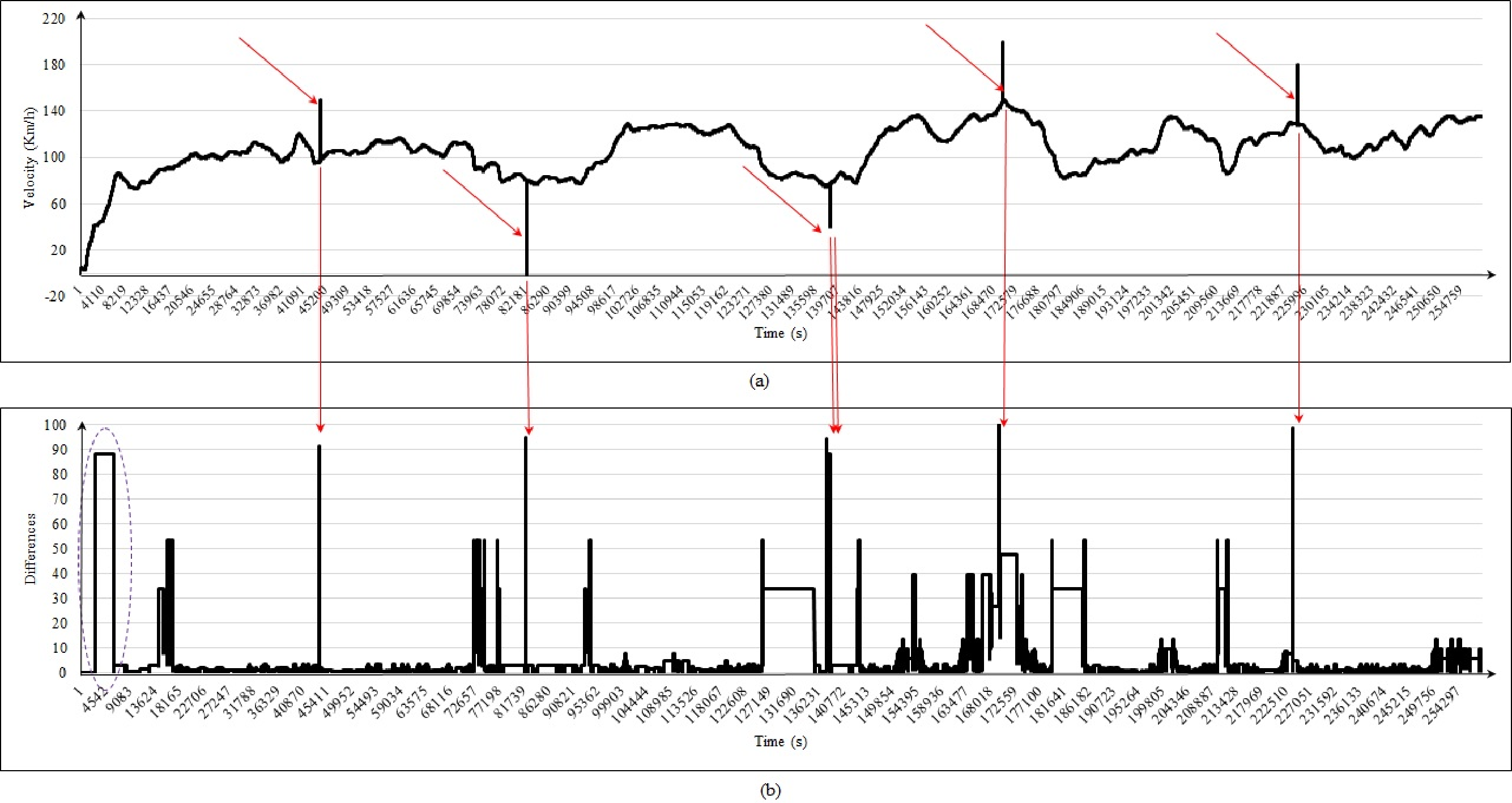}
\caption{(a) Injected anomalies under the highway scenario; (b) Detected anomalies by our proposed method under the highway scenario}
\label{fig:highway_result}
\end{figure}

\begin{figure}[t]
\centering
\includegraphics[width=3.5in]{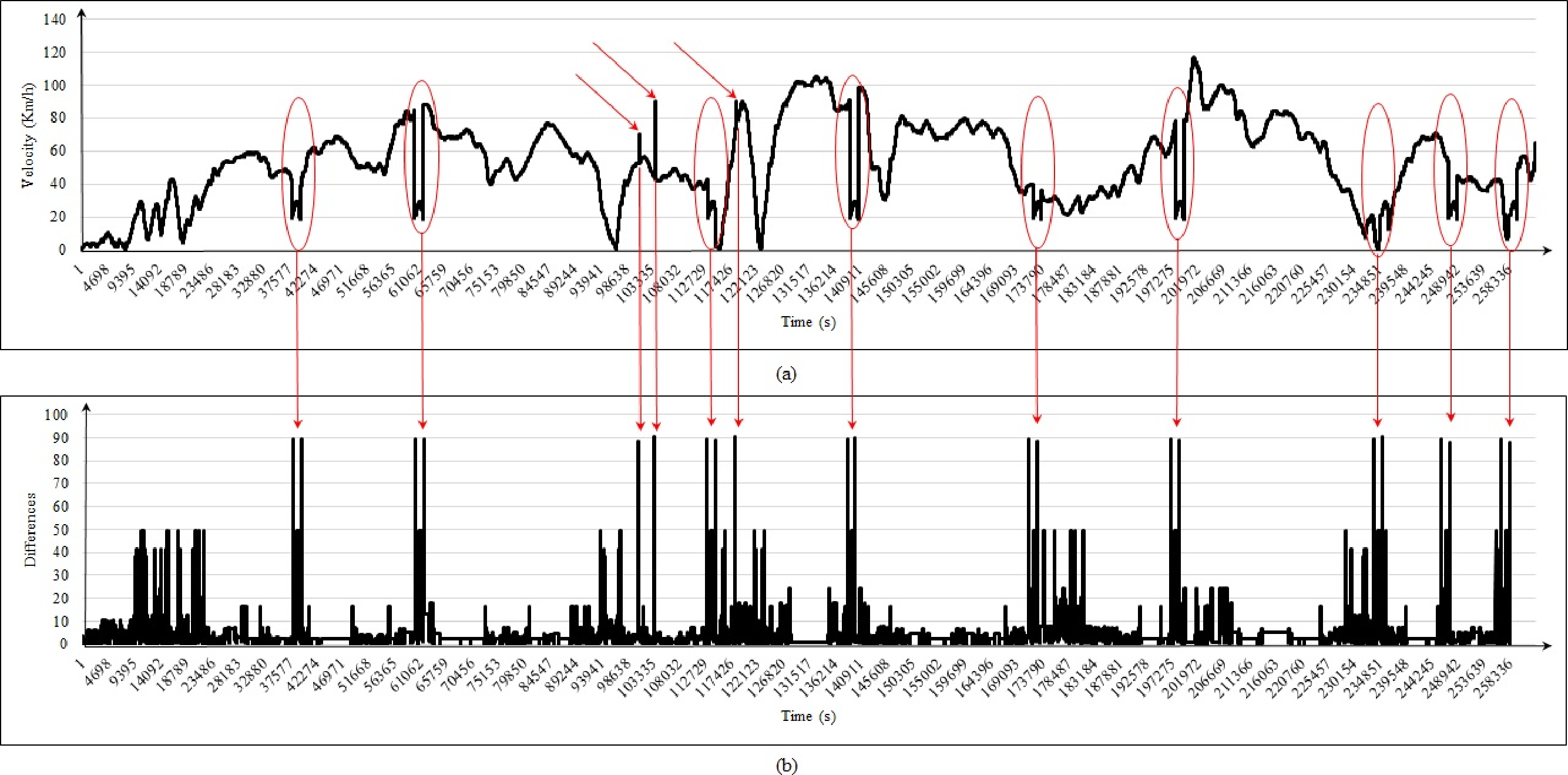}
\caption{(a) Injected anomalies under the urban scenario; (b) Detected anomalies by our proposed method under the urban scenario}
\label{fig:urban_result}
\end{figure}

We collect the velocity value of a vehicle in CAN bus and convert them into the physical value. We classify the collected datasets under the highway and the urban scenarios, as shown in Fig.~\ref{fig:highway_result} (a) and Fig.~\ref{fig:urban_result} (a), respectively. We collect 27,4487 and 26,3023 samples under the highway and the urban scenario, respectively, in which there have 260 samples for 1 second. The highway's driving behavior will be relatively smooth with less sudden braking than the one on the urban, which has more sudden acceleration. The velocity value collected from CAN bus from gateway ECU is ranging from $0$ Km/h to $160$ Km/h with around $1,000$ seconds.

The definition of anomaly is an abnormal sequence of data that does not fit with the rest of the data pattern. The proposed method focuses on a one-time attack on data payload or content, which is quite challenging among all the existing work. Here, we inject two types of attacks related to one-time: (1) Bad injection attack (one-time attack / random attack); (2) Replay attack. The first type of attack is that an attacker injects a random sequence of messages that (most likely) has never appeared before in the data history. The second type of attack is that an attacker injects a sequence of messages that have been previously read from the data. Marchetti and Stabili's work~\cite{Marchetti17} cannot detect replay attack. 

Fig.~\ref{fig:highway_result} shows the injected anomalies under the highway scenario, in which we have randomly injected anomalies with six bad injection attacks. Fig.~\ref{fig:urban_result} shows the injected anomalies under the urban scenario. We have randomly injected anomalies with three bad injection attacks and nine replay attacks. Fig.~\ref{fig:highway_result} (a) shows 6 randomly injected anomalies. Fig.~\ref{fig:highway_result} (b) shows the detection results using our proposed method, in which there have seven anomalies have been detected. The reason for the false detected anomaly (cycling with dotted dash line in purple) is that we have no big enough training dataset to cover all the scenarios. In addition, the dataset under the highway scenario quite consists of data value. The rest six anomalies are quite obvious detected in Fig.~\ref{fig:highway_result} (b) highlighted using red arrows. Our proposed method uses only $4,544,221,137$ clock cycles, which is around $1.748$ second to construct the detest matrix and provide the results.

Fig.~\ref{fig:urban_result} shows the collected dataset under the urban scenario and results of anomalies detection. Fig.~\ref{fig:urban_result} (b) shows the detection results using our proposed method, in which there has 12 anomalies have been detected explains 12 randomly injected anomalies in details). Compared to the results under the highway scenario, the result under the urban scenario reaches a 100\% detection rate and 0\% false-positive rate. Our proposed method uses $4,708,523,376$ clock cycles, which is around $1.811$ second to construct the detest matrix and provide the final results under the urban scenario. 

\begin{table}[h]
\renewcommand{\arraystretch}{1}
\caption{Comparisons of the proposed method with the previous methods to detect ``one-time" attacks}
\label{table:final}
\centering
\begin{center}
\begin{tabular}{|l|c|c|c|c|c|c|}\hline
Techniques    & False  & Anom. & FPR& Train & Test & Codesize  \\
  & Anom. & & & (s)  & (s)   &(Kbytes)\\ \hline
SVDD*~\cite{Theissler14} & 10  & 51      &16.4\%  & 0.23     & 0.25 &274   \\\hline
HMM*~\cite{Narayanan15}  & 6   & 10     &37.5\%   & 0.33     & 0.32  &218  \\\hline
LSTM*~\cite{Chockalingam17} & 4   & 42     &8.7\%   & 981     & 4.7    &20\\\hline
Our proposed & 2   & 201       &\textbf{1\%}& 0.13     & \textbf{0.15}   & \textbf{4.1}\\\hline
\end{tabular}
\end{center}
FPR: False Positive Rate; Anom. Anomalies *: state-of-the-art open-source available models for anomaly detection in-vehicle network
\end{table}

Moreover, we also test the results of the false-positive rate (FPR) under different injected anomalies. We inject different anomalies by increasing their value ranging among 10\%, 20\%, and 40\%. We also reconstruct the training model with the original dataset + hacked dataset and hacked dataset + original dataset as shown in Table~\ref{table:final}. From Table~\ref{table:final},  FPRs for case 4 are 50\%, which implies that the noise from the urban datasets may have affected the prediction of the learning process. However, FPR is successfully reduced to 41\% with a 40\% anomaly deviation. It shows that our proposed real-time ADS is robust against the noise of the training and testing datasets.

Finally, we also evaluate the performance of the existing methods and our proposed method under the ``one-time" attacks anomalies. To the best of our knowledge,  the state-of-the-art open source models for anomaly detection in-vehicle network are HMM based ADS~\cite{Narayanan15}, SVDD based ADS~\cite{Theissler14}, and LSTM based ADS~\cite{Chockalingam17}. In order to have a fair comparison, we use open-source codes with our pre-processed dataset to be fed into the models above. The data pre-processing for each model has been detailed in Section~\ref{section:result} Experimental Setup. Note here, it is the first time to compare the performance of anomalies detection among different machine learning methods using the same collected dataset under the same hacking scenario (``one-time" attack). Table~\ref{table:final} shows the comparison between our proposed method with the existing models under the ``one-time" attacks. Experimental results show that the proposed method could achieve 0.13 seconds for training and 0.15 for testing, respectively, which is 54.4\%,  40\%, and 96.8\% faster than existing HMM, SVDD, and LSTM models based ADS. In addition, our proposed ADS occupies the smallest code size (4.1 Kbytes) with a 1\% false-positive rate (FPR) to detect the challenging anomalies (one-time attacks and replay attacks) among the existing methods in-vehicle network.


\section{Conclusion}
\label{section:conclusion}
In this paper, in order to prevent ``one-time" and replay attacks to CAN bus in-vehicle network, we have proposed a novel real-time ADS based on the unsupervised machine learning method, which ensures that the method does not need training data for anomalies or attacks. We adopt self-information of information theory to construct the values of training and testing matrices, leading to successfully detect the challenging attacks in-vehicle network, e.g., ``one-time" and replay attacks. It can be easily extended to detect message flooding and cyclic message attacks. The dataset is directly collected from CAN bus containing more practical CAN bus information than the one collected from the OBD-II port, in which the collection rate can be faster than the CAN bus's baud rate. Then, we have performed verification of training and testing under the highway and urban scenarios, which shows that the FPR decreases if injecting with a larger value of an anomaly. Finally, we also use open source codes to implement HMM, SVDD, and LSTM based ADS to achieve a fair comparison with proper pre-processing of our collected dataset. Compared to the best among the state-of-the-art models, e.g., HMM, SVDD, and STM based detection, experimental results show that our proposed method achieves 8.7 times lower FPR, 1.77 times faster testing time, and 4.88 times smaller footprint. 


\end{document}